\documentclass{emulateapj}

\usepackage{apjfonts}

\newcommand{\tnm}{\tablenotemark}

\newcommand{\flux}{ergs cm$^{-2}$ s$^{-1}$}
\newcommand{\intens}{ergs cm$^{-2}$ s$^{-1}$ deg$^{-2}$}

\newcommand{\hst}{{\it HST}}

\newcommand{\chandra}{{\it Chandra}}
\newcommand{\spitzer}{{\it Spitzer}}

\newcommand{\logn}{$\log{N}/\log{S}$}

\newcommand{\W}{\hphantom{0}}
\newcommand{\Wp}{\hphantom{.}}
\newcommand{\Wf}{\hphantom{\footnotemark{a}}}

\newcommand{\z}{$z_{850}$}
\newcommand{\B}{$B_{435}$}
\newcommand{\V}{$V_{606}$}
\newcommand{\iband}{$i_{775}$}
\newcommand{\miz}{$i_{775}-z_{850}$}
\newcommand{\mBV}{$B_{435}-V_{606}$}

\newcommand{\hm}{\citetalias{hick06a}}
\newcommand{\hmp}{\citepalias{hick06a}}

\begin{document}

\title{Resolving the unresolved cosmic X-ray background in the {\it
    Chandra} Deep Fields}
\shorttitle{RESOLVING THE COSMIC X-RAY BACKGROUND}
\shortauthors{HICKOX \& MARKEVITCH}
\author{Ryan C. Hickox\altaffilmark{1}}
\author{Maxim Markevitch\altaffilmark{1}}
\altaffiltext{1}{Harvard-Smithsonian Center for Astrophysics, 60 Garden Street,
 Cambridge, MA 02138; rhickox@cfa.harvard.edu, maxim@head.cfa.harvard.edu.}

\slugcomment{Accepted for publication in {\it Astrophysical Journal Letters}}

\begin{abstract}
We present a measurement of the surface brightness of the cosmic X-ray
background (CXB) in the \chandra\ Deep Fields, after excluding all
detected X-ray, optical and infrared sources.  The work is motivated
by a recent X-ray stacking analysis by Worsley and collaborators,
which showed that galaxies detected by \hst\ but not by \chandra\ may
account for most of the unresolved CXB at $E>1$ keV.  We find that
after excluding \hst\ and \spitzer\ IRAC sources, some CXB still
remains, but it is marginally significant: $(3.4\pm1.4)\times10^{-13}$
\intens\ in the 1--2 keV band and $(4\pm9)\times10^{-13}$ \intens\ in
the 2--5 keV band, or $7\%\pm3\%$ and $4\%\pm9\%$ of the total CXB,
respectively.  Of the 1--2 keV signal resolved by the \hst\ sources,
$34\%\pm 2\%$ comes from objects with optical colors typical of
``normal'' galaxies (which make up 25\% of the \hst\ sources), while
the remaining flux comes from
objects with colors of starburst and irregular galaxies. In the 0.65--1 keV band (just above the bright Galactic
\ion{O}{7} line) the remaining diffuse intensity is
$(1.0\pm0.2)\times10^{-12}$ \intens.  This flux includes emission
from the Galaxy as well as from the hypothetical warm-hot
intergalactic medium (WHIM), and provides a conservative upper limit
on the WHIM signal that comes interestingly close to theoretical
predictions.

\end{abstract}

\defcitealias{hick06a}{HM06}

\section{Introduction}
The cosmic X-ray background (CXB) is known to be primarily
the integrated emission from X-ray point sources,
mostly active galactic nuclei \citep[e.g.,][]{hasi05,baue04}.  Recent studies
have shown that the \chandra\ Deep Fields (CDFs) North and South, the
most sensitive observations of the X-ray sky to date, resolve $\sim$80\%
of the extragalactic CXB at 1--2 keV \citep[][hereafter
HM06]{more03, wors05,hick06a}, and
$\sim$50\% at 7 keV \citep{wors05}.  While the majority of the CXB at
$E<5$ keV has been accounted for, the nature of the still-unresolved
component (the absolute intensity of which we derived in \hm) remains unclear.

Part of the unresolved CXB is due to faint point
sources that have yet to be detected in X-rays.  However, \hm\ showed
that an extrapolation of the observed \logn\ curve to zero fluxes
falls far short of the unresolved CXB.  To account for it, the slope of the
relation must steepen below the CDF-N X-ray source detection limit of
$S_{\rm 0.5-2\; \rm keV}= 2.4\times10^{-17}$ \flux.  This is not
implausible; \citet{baue04} showed that the CDFs contain a population
of normal and starburst galaxies with $S_{\rm 0.5-2\; \rm keV}
<10^{-16}$ \flux\ whose \logn\ is indeed steeper, and which should
dominate the source counts at $S_{\rm 0.5-2\; \rm keV}<10^{-17}$
\flux.  These faint sources, while not detected in X-rays, can be
associated with optical or infrared sources found in the
 CDFs.  A recent X-ray stacking
analysis of {\it Hubble Space Telescope} (\hst) sources in the CDFs
has found that their X-ray emission can account for most of the
unresolved $E=1-6$ keV CXB \citep{wors06}.

Still, some of the unresolved flux should be truly diffuse.  At $E<1$
keV, there is flux from the Local
Bubble \citep{snow04} as well as charge exchange local to the Sun
\citep[e.g.,][]{crav00, warg04}.    In addition, some signal may come
from the warm-hot intergalactic medium (WHIM) with temperatures
$10^5$--$10^7$ K, that is thought to comprise the majority of the
baryons in the local Universe
\citep[e.g.,][]{cen99,dave01}.  Models of the WHIM predict a signal
dominated by line emission from oxygen and iron with prominent lines
at 0.5--0.8 keV, but some flux up to 1.5 keV \citep{phil01,fang05a,
ursi06}.  To date the WHIM has proved challenging to observe, although
there have been reports of detections of WHIM X-ray absorption lines
\citep[e.g,][however see Kaastra et~al.\ 2006]{nica05}, as well as
excess soft X-ray emission around galaxies that has been attributed to the
WHIM \citep[e.g.,][]{solt06}.  Diffuse emission could
also come from more exotic sources such as decay of dark matter particles
\citep[e.g.,][]{abaz07}.

\nocite{kaas06}

 In this Letter, we constrain the diffuse CXB by directly measuring
its absolute intensity in deep \chandra\ pointings after the removal
of all sources detected in deep X-ray, optical, and IR observations.
Our analysis is complementary to that of \citet{wors06}, and is a
follow-up to \citetalias{hick06a}, in which we excluded only the X-ray
sources.

\section{Data}
\label{data}
{\it Chandra} X-ray observations of the CDFs North and South (CDF-N
and CDF-S) have respective total exposure times of roughly 2 Ms and 1
Ms.  The CDF-N data consist of two subsets, observed in in Very Faint
(VF) or Faint (F) telemetry mode.  Each of the three subsets (CDF-N
VF, CDF-N F, and CDF-S) contains 6 to 9 individual observations with
exposure times 30 to 170 ks each.  The datasets are identical to those
used in \hm, except for the ACIS calibration updates.  For subtraction
of the instrumental background, we add the latest ACIS stowed
observations, for a total background exposure of 325 ks, compared to
256 ks used in \hm.  These  updates caused no significant change to the
CXB fluxes.

In \hm, we measured the CXB intensity after excluding X-ray
detected sources from the catalog of \citet{alex03}. To further
exclude the contributions of X-ray faint objects, we use sources detected in the optical and
IR observations of the Great Observatories Origins Deep Survey
\citep[GOODS,][]{dick03}.  Optical and near-IR data were taken using
the Advanced Camera for Surveys (ACS) on \hst\ in the \B, \V,
\iband, and \z\ bands \citep{giav04}.  We use the public
catalog\footnotemark\ based on detections in the \z\ band ($\lambda \sim 8300-9500$ \AA), with a limiting magnitude
of $z_{850}\simeq 27$.  The \hst\ catalogs for CDF-N and CDF-S
each contain $\sim$4500 objects within the area for which we extract CXB
spectra (inside a 3.2\arcmin\ circle, and away from detected X-ray
sources, see \S\ \ref{analysis}).

\footnotetext{http://archive.stsci.edu/pub/hlsp/goods/catalog\_r1/h\_r1.1z\_readme.html}

We also exclude sources detected with the Infrared Array Camera (IRAC)
on {\it Spitzer}, in four bands centered on 3.6, 4.5, 5.8, and 8
$\mu$m.  There is no published catalog currently available for these
sources, so we have performed our own source detection using the
calibrated IRAC images\footnotemark.  Sources were detected using the
{\tt SExtractor} code \citep{bert96} in all four bands.  In the most
sensitive band (3.6 $\mu$m), we detect $\sim$1000 sources in the X-ray
extraction region to a 5$\sigma$ limiting flux of $\sim$0.2 $\mu$Jy.
Most ($>$95\%) of the IRAC sources have counterparts within 1\arcsec\
in the \hst\ \z\ catalog.

\footnotetext{http://data.spitzer.caltech.edu/popular/goods/Documents\\/goods\_dataproducts.html}
\setcounter{footnote}{1} 

We also considered \spitzer\ Multiband Imaging
Photometer (MIPS) 24 $\mu$m sources\footnotemark\ and for CDF-N, Very
Large Array (VLA) 21 cm sources \citep{rich00}, with limiting fluxes
of 80 and 40 $\mu$Jy, respectively.  All the sources within our
X-ray extraction regions have \hst\ or IRAC counterparts, except
for 2--4 of the 80--90 MIPS sources and 1 of the 6 VLA sources, so we
do not include these data in the analysis.

\begin{figure}
\plotone{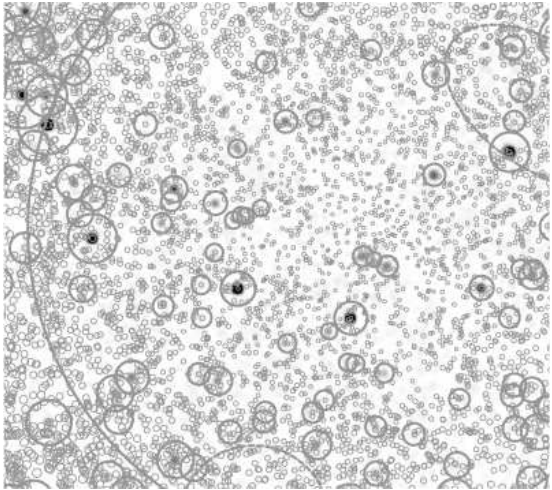}
\caption{Smoothed 0.5--8 keV image of CDF-N, showing the source exclusion
  regions.  The large circle is the 3.2\arcmin\ extraction region,
  medium-size circles and ellipses show X-ray point and extended
  sources, and small circles show \hst\ sources. \label{figim}}
\end{figure}

\section{Analysis}
\label{analysis}
We calculate the CXB surface brightness as discussed in detail in \S\
5 of \hm, with only minor differences.  For the \hst\ and IRAC sources
not detected in X-rays, we use smaller exclusion regions with radii
equal to  $r_{90}$, the X-ray point-spread function (PSF) 90\% enclosed flux
radius at 1.5 keV (see Eqn.\ 1 of \hm), compared to 4.5--$9r_{90}$ for
X-ray detected sources.  Due to the large number of optical and IR
sources, at large off-axis angles there is little area remaining after
source exclusion.  Therefore we only extract spectra from a circle
around the optical axis with radius 3.2\arcmin\ (where
$r_{90}=2.2$\arcsec), compared to 5\arcmin\ in \hm.  The solid angle
we use here is 0.45 of that in \hm\ when we exclude only X-ray
sources, and 0.32 when we exclude all \hst\ and IRAC sources.

\begin{figure}
\epsscale{1.15}
\plotone{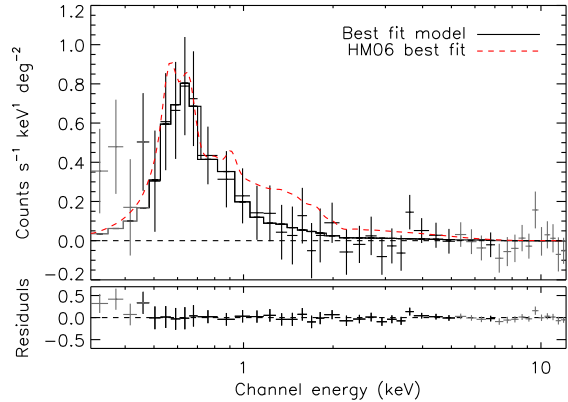}
\caption{CXB spectrum for the CDF-N VF subset, with
  all sources (\chandra, \hst\, and IRAC) excluded.  The spectrum is
  fit with an APEC model with $kT=0.20$ keV plus a power law with
  $\Gamma=1.5$. Black points show the data included in the fit (see \hm), and
  the red dashed line shows the best-fit model for the CDF-N VF
  spectrum excluding only X-ray sources  \hmp. \label{figspecvf}}
\end{figure}

For each of the datasets, we fit the spectrum of the remaining area
with a model consisting of an optically-thin thermal component (APEC)
with solar abundances and $kT\simeq 0.15$--0.2 keV, to account for
emission from the Galaxy as well as any diffuse thermal CXB.  We also
include a power law component that dominates at $E>1$ keV, and fix the
photon index at $\Gamma=1.5$, the best-fit value from \hm\ (we do not
vary $\Gamma$ as in \hm, due to poorer statistics).  These models do
not necessarily provide a physical description of the emission, but
they are sufficient for our purpose of calculating fluxes.  A CDF-N VF
spectrum with the model fit is shown in Fig.\ \ref{figspecvf}.

\begin{deluxetable}{lcccc}
\tablecaption{CXB surface brightness after excluding all sources}
\tablewidth{3in}
\tablehead{
\colhead{Energy} &
\multicolumn{3}{c}{CXB intensity\tnm{a}} &
\colhead{CXB frac.\tnm{b}} \\
\colhead{(keV)} &
\colhead{CDF-S} &
\colhead{CDF-N (VF+F)} &
\colhead{Average} &
\colhead{(\%)}}
\startdata
0.5--2\tnm{c} & $45\pm 8$\W & \Wf$41\pm 11$\tnm{d}\W\Wp & \Wf$43\pm 7$\tnm{d}\W\Wp &  \nodata \\
0.65--1\tnm{c} & $11\pm 2$\W & \Wf$9\pm 3$\tnm{d}\Wp & \Wf$10\pm 2$\tnm{d}\W\Wp &  \nodata \\
1--2 & $   5.0\pm   1.8$ & $   2.4\pm   1.6$ & $   3.4\pm   1.4$ & $   7.3\pm   3.0$ \\
1--1.3 & $   1.7\pm   0.7$ & $   1.0\pm   0.7$ & $   1.3\pm   0.6$ & $   8.3\pm   3.8$ \\
2--5 & \W$ 7\pm12$ & \W$ 2\pm10$ & $ 4\pm 9$ & $ 4\pm 9$ \\
2--8 & $14\pm39$ & \W$ 4\pm30$ & \W$ 7\pm26$ & \W$ 4\pm16$ 
\enddata
\tablenotetext{a}{Units are $10^{-13}$ \intens, errors are 68\%.}
\tablenotetext{b}{Percent of total CXB, assuming a
power-law CXB spectrum with $\Gamma=1.4$ and normalization $10.9\pm0.5$ photons cm$^{-2}$
s$^{-1}$ sr$^{-1}$ keV$^{-1}$ at 1 keV (\hm).}
\tablenotetext{c}{Intensities for $E<1$ keV are dominated by Galactic
and local line emission that varies with time and position.  These are
not extragalactic values.}
\tablenotetext{d}{Does not include CDF-N VF, to exclude the local temporary excess.}
\end{deluxetable}

\begin{figure*}
\plottwo{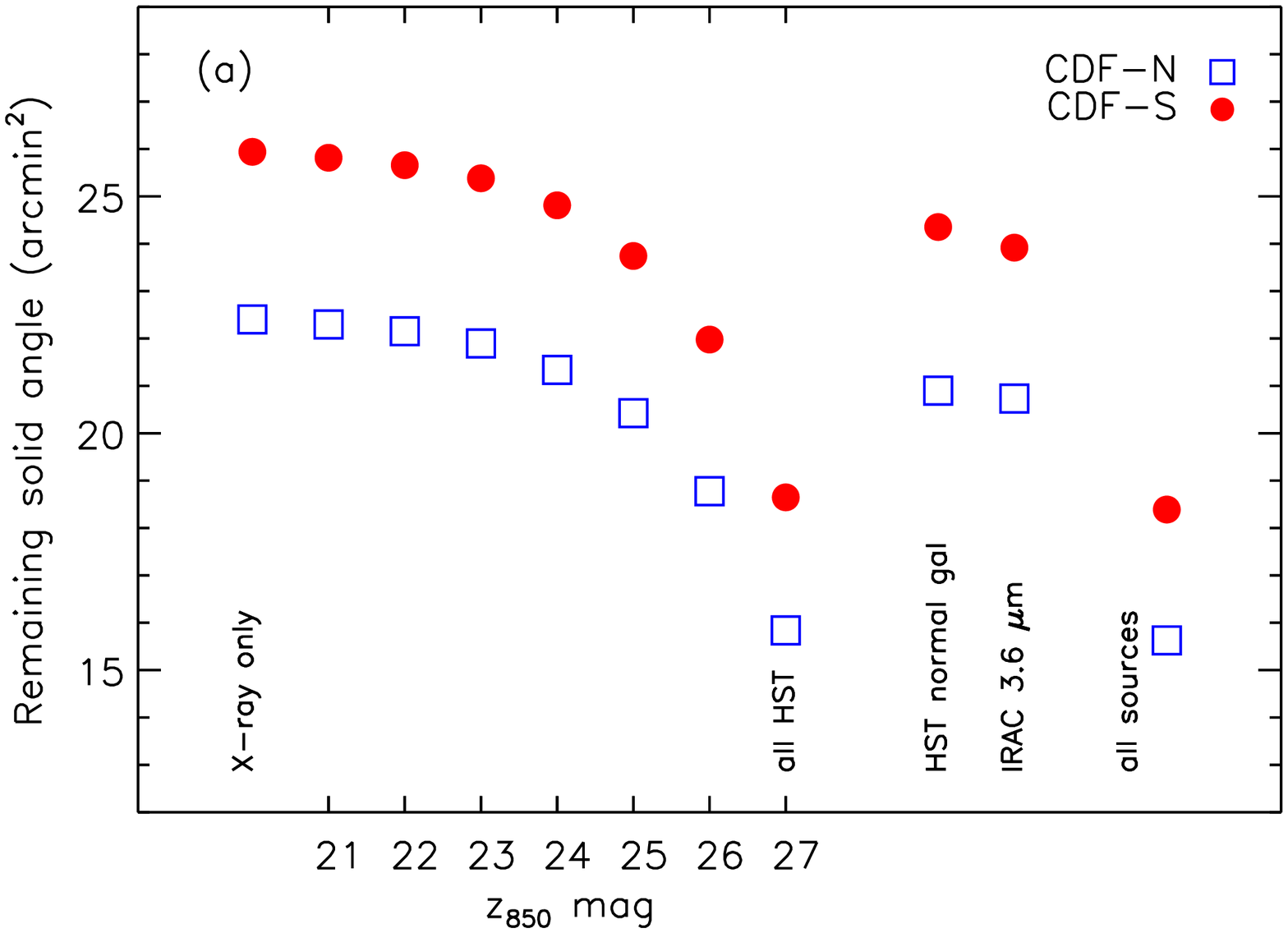}{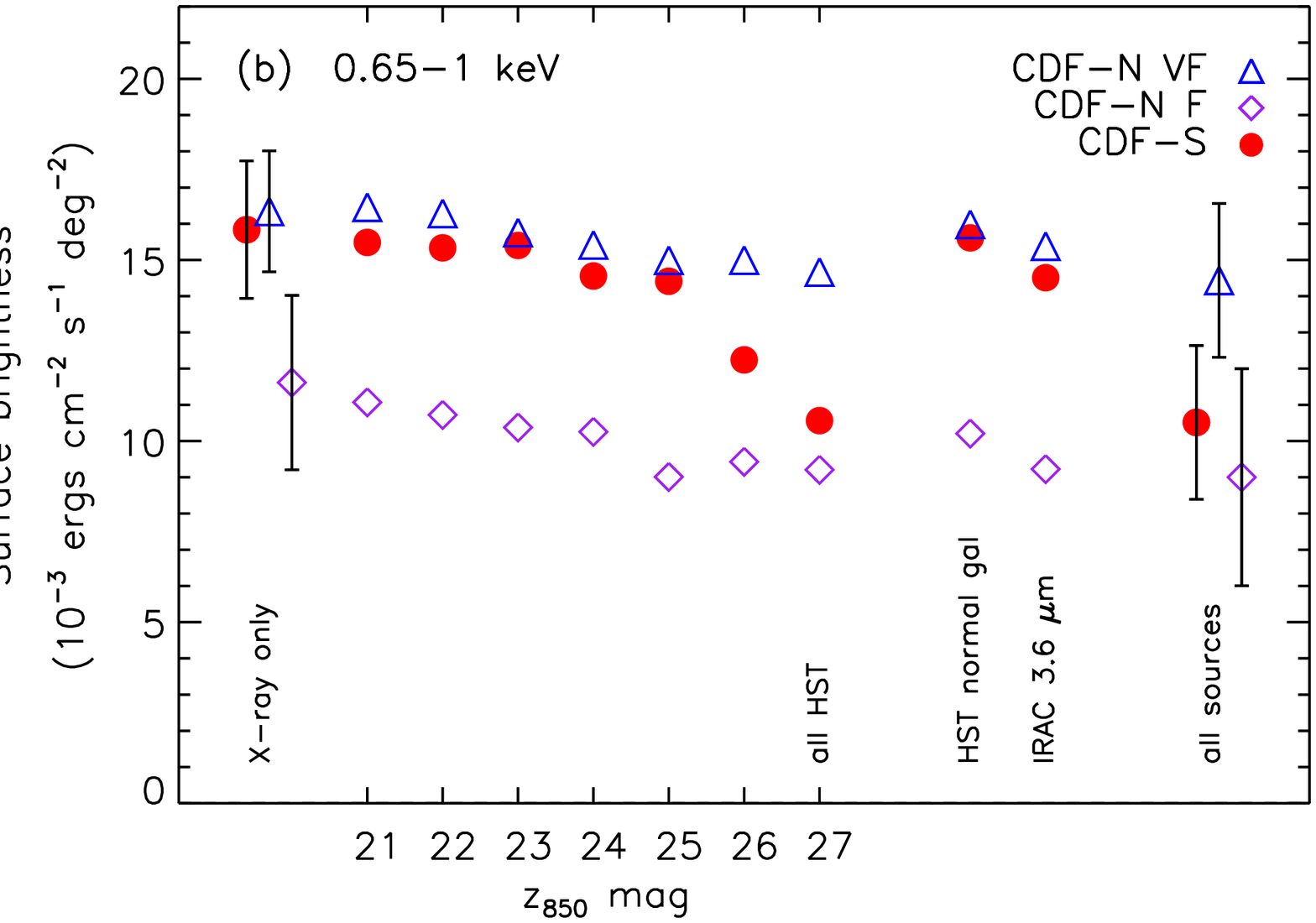}
\vskip-0.3cm
\plottwo{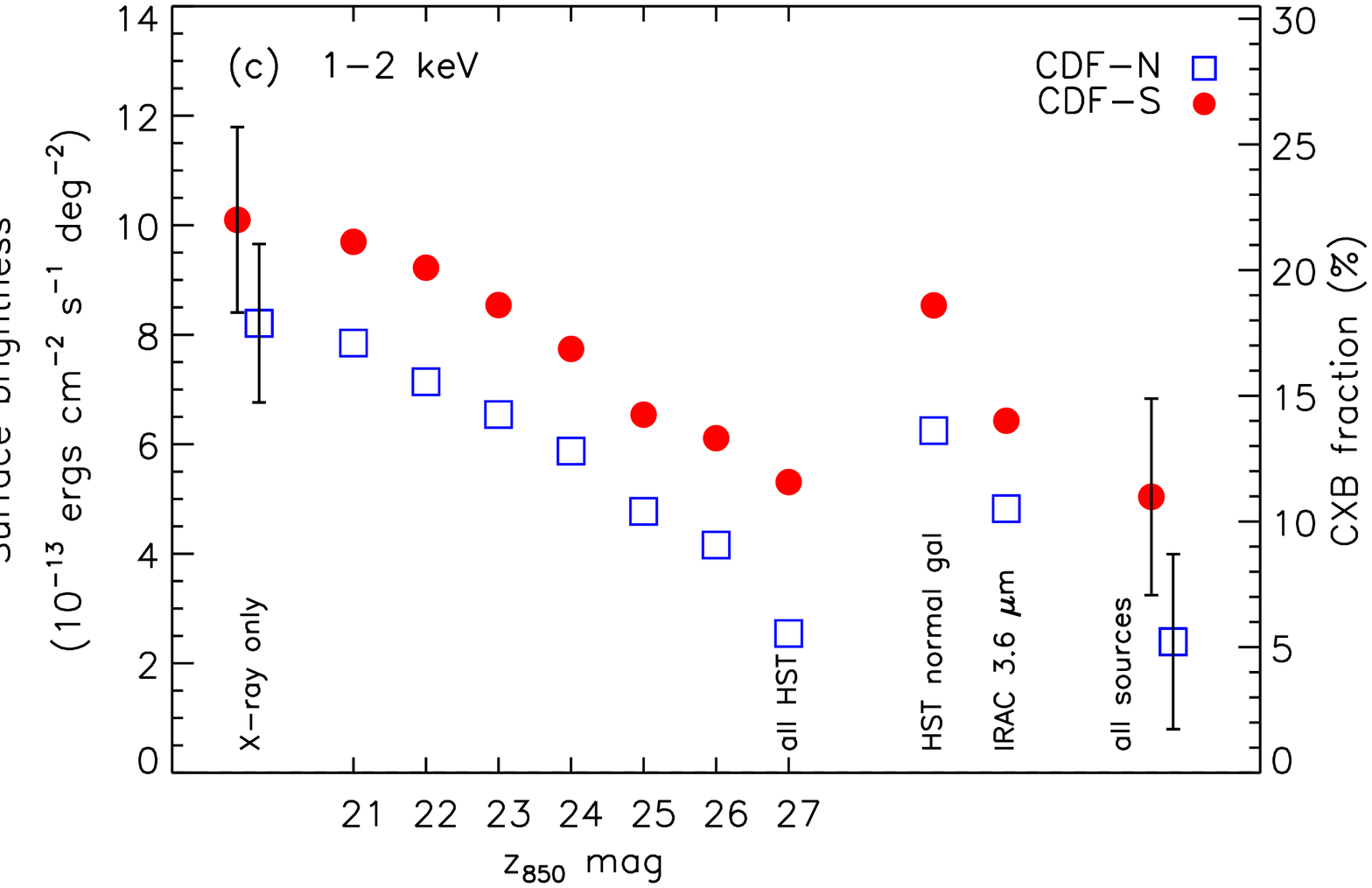}{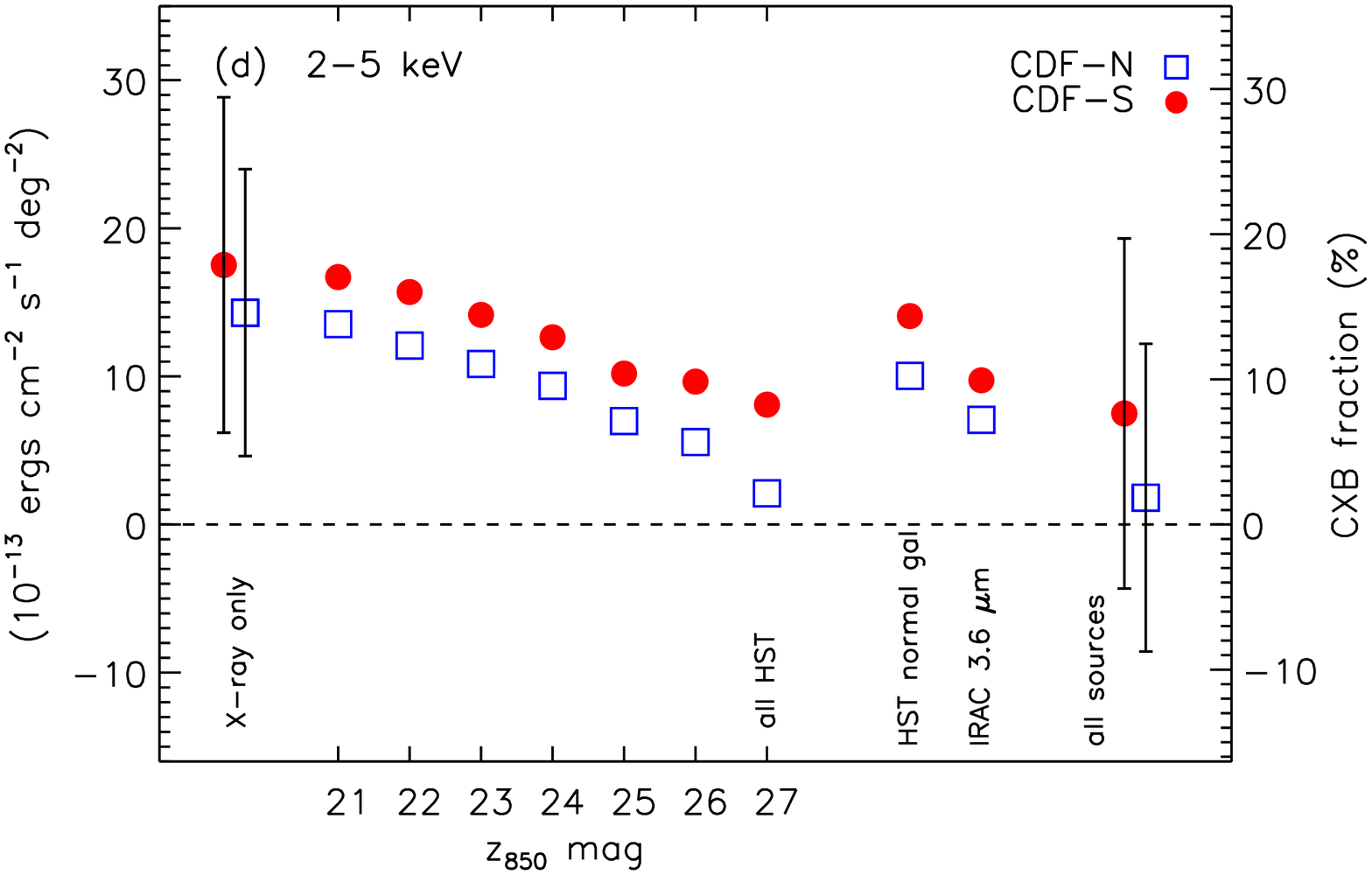}
\caption{(a) Solid angles of the X-ray extraction regions, and (b--d)
  unresolved CXB surface brightness, for subsets of excluded \chandra,
  \hst, and IRAC sources.  In all panels, points on the far left show
  the value when only X-ray sources are excluded (as in \hm, but
  using the smaller 3.2 \arcmin\ region).  Points with \z\ labels on
  the $x$-axis
  correspond to the additional exclusion of \hst\ sources brighter
  than \z\ (note that $z_{850}=27$ is only an approximate detection
  limit for all HST sources).  Points on the right show intensities
  excluding X-ray sources plus only \hst\ ``normal'' galaxies (\S\
  \ref{whichsource}), only IRAC 3.6 $\mu$m sources, and finally all \hst\ and
  IRAC sources.  For the 0.65--1 keV band we show the CDF-N VF and
  CDF-N F subsets separately, as time variability in the local
  emission \citep[apparently charge exchange in the Solar wind,
  e.g.,][]{warg04} caused these intensities to differ.  For clarity,
  errors are shown only on the far left and right points.  For (c) and
  (d), on the right axis we show the corresponding fraction of the
  total CXB, for which we assume a power law
  spectrum with $\Gamma=1.4$ and normalization 10.9 photons cm$^{-2}$
  s$^{-1}$ sr$^{-1}$ at 1 keV \hmp. \label{figres}}
\end{figure*}

For each spectrum, we calculate fluxes in the 0.5--2 keV, 1--2
keV, 2--8 keV, and 2--5 keV bands, as well as the 0.65--1 keV and 1--1.3 keV
bands, where the hypothetical WHIM signal or galaxy group emission are
best distinguished from the diffuse Galactic foreground (see \S\
\ref{whim}).  Errors on the observed flux include statistical errors
in the sky and background count rates, and a 2\% systematic
uncertainty in the background normalization (\S\ 4 of \hm).  We
carefully account for background uncertainties.  Because of
differences in pointing of up to 2\arcmin\ and the small X-ray
extraction area, the composite stowed background spectrum includes 3
times (for CDF-N) and 1.5 times (for CDF-S) as many statistically
independent photons as do the spectra for each separate observation.
We accordingly increase the effective background exposure times by
these factors.  We conservatively neglected this effect in
HM06, as it was much smaller due to the larger extraction area and
fewer excluded sources.

A further small correction is necessary to subtract the point source
flux that is scattered outside our exclusion regions.  For each \hst\
or IRAC source, 10\% of the X-ray flux should lie outside $r_{90}$.
However, because many regions overlap, the
contribution of this scattered flux is smaller.  Using a model of the
\chandra\ PSF, we estimate that when all \hst\ and IRAC sources are
excluded with radius $r_{90}$, 5\% (0.5--2 keV) and 8\% (2--8 keV) of
their flux should remain in our unresolved spectrum.  Likewise, 0.3\%
and 0.6\% of the flux from detected
X-ray sources (for which we use larger exclusion regions) will
be scattered into the spectrum.  We account for both these small
contributions in the final unresolved intensities.  As a check, we
tried exclusion regions of radius $1.5r_{90}$ and obtained
identical results.

\section{Results}
We first calculate the CXB intensity after excluding \hst\ subsamples with various limiting magnitudes from
$z_{850}<21-26$, and finally excluding all \hst\ sources
($z\lesssim27$).  We also try excluding only ``normal'' galaxies, selected
based on their \hst\ colors (see \S\ \ref{whichsource}), only  IRAC
3.6 $\mu$m sources, and finally all \hst\ and IRAC sources.  In every
case we also exclude all detected X-ray sources.  The results are
shown in Fig.\ \ref{figres}, and the residual CXB intensities excluding
all sources are given in Table
1.  We also calculate the mean CXB intensity for all three subsets, including a detailed error
analysis as described in \S\ 8.2.3 of \hm.

Fig.\ \ref{figres} shows that excluding faint \hst\ sources has little
effect on the CXB at $E<1$ keV, where it is dominated by the truly
diffuse Galactic and local emission.  At $E>1$ keV, as fainter objects
are excluded, the surface brightness of the CXB decreases, indicating
that the \hst\ sources do indeed contribute to the X-ray unresolved
CXB.  In the 1--2 keV band, the remaining intensity after excluding
all \hst\ and IRAC sources is above zero at only 1.5$\sigma$ for CDF-N
and 2.8$\sigma$ for CDF-S.  The CXB intensity decreases even as we
excluded the faintest \hst\ sources, suggesting that if we were able
to go slightly deeper, we would resolve the 1--2 keV CXB to within our
absolute uncertainties.  The X-ray undetected \hst\ sources contribute
1--2 keV intensities of $(5.7\pm0.3)\times10^{-13}$ and
$(4.3\pm0.3)\times10^{-13}$ \intens\ in CDF-N and CDF-S, respectively,
or  $59\%\pm9\%$ and
$45\%\pm7\%$ of the unresolved CXB from \hm.
The fluxes resolved by \hst\ sources (given here and in
\S~\ref{whichsource}) represent {\em differences} between the points
in Fig.~\ref{figres}, and do not depend strongly on the instrumental
background.  Thus their errors are much smaller than the uncertainties
on the absolute surface brightness shown in Fig.~\ref{figres}. 
Stacking the X-rays from the same \hst\ sources, \citet{wors06} found larger
intensities of $(8.4\pm0.5)\times10^{-13}$ and
$(9.3\pm0.7)\times10^{-13}$ \intens\ for CDF-N and CDF-S,
respectively.  The cause of this difference is not immediately clear,
but may be related to different accounting for overlap between the
\hst\ source regions.

\begin{figure}
\epsscale{1.1}
\plotone{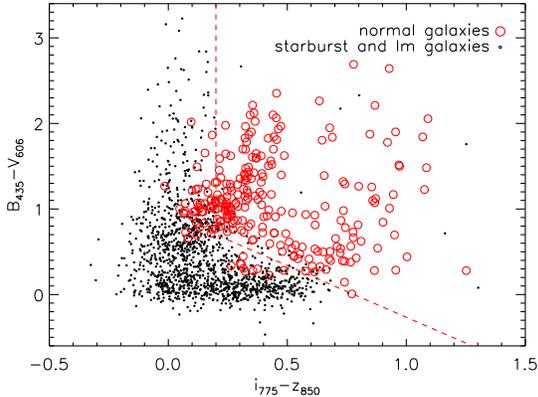}
\caption{Color selection of ``normal'' galaxies from \hst\ data.  All
  points show $B_{435}-V_{606}$ vs. $i_{775}-z_{850}$ colors for
  sources in the UDF with $z_{850}<27$.  Sources are classified as
  elliptical and spiral galaxies (red circles) or starburst and Im
  galaxies (black dots) from photo-$z$ analysis \citep{coe06}.  Lines
  show our criteria to select ``normal'' galaxies from the GOODS data.
  \label{figcol}}
\end{figure}

In the 2--5 keV band, the \hst\ and IRAC sources account for the
entire unresolved CXB within the large absolute uncertainties.
Similarly, \citet{wors06}  found that the stacked flux from
 \hst\ sources is consistent with completely resolving the 2--6
keV CXB.

\section{Discussion}
\subsection{Which sources resolve the CXB?}
\label{whichsource}

\citet{baue04} and \hm\ show that AGN likely contribute a relatively
small fraction of the CXB at X-ray fluxes below the CDF limit, where
the X-ray source population is dominated by starburst and normal
galaxies.  To determine what class of galaxies gives the largest
contribution to the unresolved CXB, we use \hst\ colors to
approximately divide the \z\ sources into two subsets by galaxy type.
While we do not have optical classifications for the \hst\ GOODS
galaxies, such a classification exists for the \hst\ Ultra Deep Field
(UDF), which covers a small region roughly 3\arcmin\
from the CDF-S aimpoint with a flux limit of $z_{850}\simeq30$.  The
UDF was observed in the same four ACS bands as the wider GOODS fields,
and \citet{coe06} performed a detailed photometric redshift analysis
of the UDF sources, which classified each galaxy with one of eight
templates (elliptical, Sab, Sbc, Im, and four types of starburst).

In Fig.\ \ref{figcol}, we show \mBV\ vs. \miz\ colors for
\citet{coe06} sources with $z_{850}<27$, corresponding to the flux
limits of the larger GOODS fields.  In this diagram, normal
(elliptical and spiral) galaxies occupy a different region from
starburst and Im galaxies, with approximate boundaries
$i_{775}-z_{850}=0.2$ and $B_{435}-V_{606}=0.9-(i_{775}-z_{850})$, as
shown in Fig. \ref{figcol}.  Of the 255 ``normal'' galaxies in the UDF
sample, 201 (73\%) are in this region, while only 52 (21\%) of the 253
galaxies in this region have starburst or Im classifications and would
be mis-classified by this criterion.  In the GOODS catalog,
these criteria classify 25\% of  $z_{850}$ sources as ``normal'' galaxies and
70\% as ``starburst/Im'' objects (5\% of the
sources do not have detections in all four ACS bands).  The ``normal''
galaxies contribute $34\%\pm 2\%$ of the 1--2 keV intensity
resolved by all the \hst\ sources, and so have higher average
X-ray fluxes than other X-ray undetected \hst\ objects.

We also examine the contribution from the 3.6
$\mu$m sources, which are on average only 0.25 magnitudes brighter in \z\
than the \hst\ sources, but account for a disproportionately large
fraction of the unresolved CXB.  The $\sim$1000 3.6 $\mu$m sources
contribute $63\%\pm2\%$ of the total 1--2 keV CXB resolved by
all \hst\ and IRAC sources.  In comparison, the 1000 brightest \hst\
sources (which have $z_{850}<24.5$, and of which only 20\% are
IRAC sources) account for $53\%\pm2\%$.  Therefore,
the X-ray flux from undetected X-ray sources is apparently more highly
correlated with 3.6 $\mu$m flux than with \z\ flux.  This may be an
important clue to the nature of these faint X-ray sources, although
a detailed discussion is beyond the scope of this Letter.  In a
forthcoming paper, we will determine how bright these undetected sources can be
in the X-rays and whether \chandra\ could detect them in a reasonable
longer exposure.

\subsection{Comparison to WHIM predictions}

\label{whim}
While most of the remaining CXB signal should be due to Galactic or
local diffuse emission and unresolved X-ray point sources, our results
provide conservative upper limits on emission from the WHIM (Cen et
al.\ 2000 and later works).  The brightest WHIM emission feature
should be a redshifted \ion{O}{7} 0.57 keV line, but it cannot be
distinguished from the bright Galactic \ion{O}{7} line without much
better ($\sim$1 eV) spectral resolution.  However, the hotter WHIM
phases with $kT\sim 0.2-0.5$ keV would also emit significant
\ion{Fe}{17} lines at $E\approx 0.72-0.83$ keV (rest-frame). Thus, the ratio of
WHIM emission to Galactic foreground (which is well-approximated by a
$kT\approx 0.15-0.2$ keV thermal spectrum, Fig.\ 1) may be higher at
$E\gtrsim 0.6$ keV.

We can compare our 0.65--1 keV signal to predictions from cosmological
simulations.  \citet{fang05a} find WHIM intensities of $\sim3\times
10^{-8}$, $3\times 10^{-10}$, and $3\times 10^{-14}$ \intens\ toward a
galaxy group, filament, and void, respectively.  \citet{ronc06a}
obtained a sky-averaged 0.65--1 keV signal of $5\times10^{-13}$
\intens\ after excluding galaxy groups and clusters. (For these two
works, we have approximately converted fluxes from 0.1--1 keV and
0.5--2 keV to 0.65--1 keV).  The CDFs were selected to lie away from
bright groups, but if they do not lie in voids, then our observed
signal, $(1.0\pm0.2)\times10^{-12}$ \intens, comes interestingly close to the
above predictions.  Galactic superwinds may increase the WHIM
metallicity \citep{cen06}, so the line emission may be even higher. At
higher energies, our 1--2 keV signal is 15 times higher than the
predicted sky-averaged WHIM intensity of $0.19\times10^{-13}$ \intens\
\citep{phil01}, so does not give an interesting limit.  For a more
quantitative comparison, it will be necessary to excise the regions
around individual galaxies in the WHIM simulations similarly to our
analysis.  With such a simulation, one may be able to place
interesting constraints on the distribution and metallicity of the
WHIM in the CDFs.

\acknowledgements 

We thank J.P. Ostriker and L.A. Phillips for fruitful discussions, and
the referee, Franz Bauer, for helpful comments.  RCH was supported by
NASA GSRP and Harvard Merit Fellowships, and MM by NASA contract
NAS8-39073 and \chandra\ grant G04-5152X.

\end{document}